\documentclass[twocolumn,preprintnumbers,amsmath,amssymb]{revtex4}
\usepackage{graphicx} 
\usepackage{setspace} 
\usepackage{dcolumn} 
\usepackage{bm} 
\usepackage{natbib}
\usepackage{hyperref}
\hypersetup{colorlinks=true,linkcolor=blue,citecolor=blue,urlcolor=blue}
\usepackage{epsfig}
\usepackage{soul}
\usepackage{soul}
\usepackage{xcolor}

\begin{document}

\title{Flexible PDMS/La$_{0.7}$Sr$_{0.3}$MnO$_3$/MWCNT Composite Thin Films for Multifunctional Temperature and Magnetic Sensing Electronic Skin}

\author{Jimlee Patowary$^a$}
\author{G Suresh$^b$}
\author{Jitendra Kumar$^a$}
\author{Ashutosh Kumar$^{a,}$\footnote{Email: ashutosh@iitbhilai.ac.in}}

\affiliation{$^a$Functional Materials Laboratory, Department of Materials Science and Metallurgical Engineering, Indian Institute of Technology Bhilai, Durg 491002, Chhattisgarh, India}
\affiliation{$^b$Soft Materials Group, Department of Physics, VNR-Vignana Jyothi Institute of Engineering and Technology, Hyderabad, India}

\begin{abstract}
The development of multifunctional electronic skin (e‑skin) requires materials that combine mechanical flexibility with responsiveness to multiple stimuli. In this work, a flexible PDMS/La$_{0.7}$Sr$_{0.3}$MnO$_3$ (LSMO)/MWCNT composite thin film was fabricated via solution casting, using LSMO powder synthesized by a solid‑state reaction method. Structural and spectroscopic analyses confirm the formation of single-phase rhombohedral LSMO and successful incorporation of PDMS, LSMO, and MWCNT components. The composite exhibits a smooth and uniform surface morphology, along with significantly enhanced thermal stability, retaining $~$70\% mass at elevated temperatures. Electrical measurements reveal thermally activated resistivity behavior, enabling temperature sensing functionality. Additionally, the composite shows a notable decrease in resistance under an applied magnetic field, exhibiting magnetoresistance due to spin-dependent transport in the LSMO phase. Mechanical testing indicates elastomeric behavior with a maximum load of ~0.49 N and stretchability of $~$26\%, along with ductile deformation characteristics. The multifunctional sensing properties arise from the synergistic interaction between the conductive MWCNT network and magnetically active LSMO within the flexible PDMS matrix. Overall, the composite demonstrates a unique combination of thermal stability, mechanical flexibility, and dual sensing capability, making it a promising material for next-generation e‑skin applications.\\
\end{abstract}


\maketitle
\section{Introduction}
Skin is the outermost protective barrier of the human body, enabling interaction with the external environment. It allows humans to perceive different shapes and textures, variations in temperature, and different degrees in the intensity of contact pressure~\cite{hammock201325th}. Human skin is composed of several sensory receptors, including nociceptors for pain sensing, pruriceptors for itch sensation, thermoreceptors for temperature detection, and low-threshold mechanoreceptors (Mercel cells, Meissner and Pacinian corpuscles) responsible for sensing non-painful mechanical stimuli like touch~\cite{zimmerman2014gentle, chortos2016pursuing}. Global research initiatives are actively advancing the development of sensing systems designed to artificially reproduce the complex sensory and functional characteristics of human skin. Such artificial skins are commonly referred to as \textit{electronic skin} or \textit{e-skin}. Electronic skin has attracted significant attention due to its potential applications in various disrupting and emerging technologies~\cite{chen2021recent, yadav2025ai}. 
A fully functional e-skin generally requires multiple sensing capabilities, including pressure, temperature, strain, shear force, humidity sensing~\cite{chen2021recent}, and multifunctional sensing~\cite{guo2021recent}. The development of multifunctional e‑skin is essential for replicating the complex tactile sensing capabilities of human skin. In particular, such systems must be capable of detecting both spatially resolved and temporally varying tactile stimuli, enabling accurate perception of pressure distribution and dynamic touch events. Achieving this level of sensory functionality is critical for advancing applications in robotics, wearable healthcare devices, and human–machine interfaces~\cite{chen2021recent, zhang2022challenges}. \\
For the successful fabrication of \textit{e-skin}, several important design considerations must be incorporated to emulate the properties of natural skin. These include flexibility, low elastic modulus, and stretchability~\cite{hammock201325th}. Since human skin is continuously subjected to various mechanical deformations during body movement, artificial electronic skin must possess sufficient stretchability to accommodate such deformations~\cite{chen2021recent,yang2024electronic}, along with adequate flexibility~\cite{li2024flexible}. These flexible and stretchable skin sensors provide a promising alternative to conventional rigid wearable devices by offering superior mechanical conformity and durability under deformation. Their ability to enable continuous and reliable physiological monitoring positions them as key components in next‑generation healthcare systems aimed at long‑term, real‑time health assessment.\cite{thorat2013functionalization}\\
The stretchability and flexibility of \textit{e-skin} are primarily governed by the polymer matrix used in the composite system. Various polymers have been explored for electronic skin applications, including polyethylene terephthalate (PET)\cite{zhao2015flexible}, polyethylene naphthalate (PEN)\cite{chou2015chameleon}, polyimide (PI)\cite{zhang2023bioinspired}, polydimethylsiloxane (PDMS)\cite{schwartz2013flexible}, thermoplastic polyurethane (TPU)\cite{tian2024ultra}, styrene--ethylene--butene--styrene (SEBS)\cite{zhang2026leakage}, PVDF-TrFe/PLA\cite{suresh2026tailored}, and others~\cite{li2024flexible, mohan2025microwave}. Among these materials, PDMS has been extensively utilized because of its excellent chemical inertness, thermal stability over a wide temperature range, optical transparency, and tunable mechanical properties~\cite{hammock201325th, li2023functional, yang2025smart}. PDMS was used as a polymer matrix with ZnO as a filler for muscle activity measurement for muscle health monitoring\cite{jugade2020pdms} and it is also used to create protrusions for utilization of sandpaper as template and MXene for the constructions of micro-protrusion rough surface on PDMS film and electrically conductive pathways \cite{chen2022skin}. MXene/PDMS-based \textit{e-skin} have shown remarkable performance in motion sensing and physiological signal monitoring, underscoring the promise of flexible conductive polymer composites for next-generation wearable electronics.\cite{bai2024skin}\\
Another crucial component in the fabrication of \textit{e-skin} is the filler material, which imparts the desired sensing functionalities to the composite film. Various categories of filler materials have been investigated for electronic skin applications, including dielectric and ceramic materials~\cite{lee2021fingerpad, yin2023advanced}, carbon nanotube (CNT)-based active materials~\cite{sun2019flexible}, graphene-based active materials~\cite{iqra2022flexible}, nanowire-based active materials, and organic/polymer-based active materials~\cite{hammock201325th}. For instance, Ba$_{0.85}$Ca$_{0.15}$Zr$_{0.10}$Ti$_{0.90}$O$_3$ (BCZT) ceramics have been incorporated into a poly(vinylidene fluoride) (PVDF) matrix for energy harvesting applications~\cite{nayak2024modulating}, while sodium niobate (NaNbO$_3$) nanorods have been introduced into PVDF to enhance its $\beta$-phase content, thereby improving its suitability for powering microwatt-scale electronic devices~\cite{anand2019effect}. Furthermore, La$_{0.67}$Sr$_{0.33}$MnO$_3$ (LSMO), a multifunctional perovskite oxide exhibiting high magnetoresistive sensitivity, low noise, and excellent biocompatibility, has emerged as a promising candidate for next-generation flexible and biomedical sensing applications~\cite{vera2023high, hou2022linearly}.\\
The present study aims to optimize composite film composition to retain mechanical flexibility and introduce thermal sensing capability into the composite film along with magnetic sensing capability. With this aim, LSMO has been selected as the ceramic filler material, and its behavior upon incorporation into the PDMS polymer matrix has been systematically investigated. The rationale for selecting LSMO lies in its excellent magnetic and electrically conductive properties\cite{salamon2001physics}. However, increasing the concentration of LSMO adversely affects the flexibility of the composite film. Therefore, maintaining an optimized balance between LSMO and PDMS is essential to preserve the mechanical flexibility of the film. To address this issue, multi-walled carbon nanotubes (MWCNTs) were introduced as an additional conductive filler to enhance the electrical conductivity while allowing a higher proportion of PDMS to be retained for improved flexibility. The incorporation of MWCNTs significantly improved the electrical conductivity of the composite film. After extensive compositional optimization, it was found that the PDMS:LSMO:MWCNT ratio of 10:9:1 provides the best overall performance for achieving both temperature sensing and magnetic/electrical sensing capabilities while maintaining mechanical flexibility.\\
\section{Experimental Details}
\subsection{Materials}
Multi-walled carbon nanotubes (MWCNTs) purchased from SHILPENT were used as the conductive nanofiller in the present study. Lanthanum oxide (LOBA, 99.9\%), strontium carbonate (LOBA, 98\%), manganese(IV) oxide (Sigma, 99\%), and ethanol (CSS, 99.9\%) were used for the synthesis of La$_{0.7}$Sr$_{0.3}$MnO$_3$ (LSMO). Silicone elastomer base (SYLGARD$^{\mathrm{TM}}$ 184) and silicone elastomer curing agent (SYLGARD$^{\mathrm{TM}}$ 184) supplied by Dow Chemical were used as the polymer matrix materials. Isopropyl alcohol (IPA) (LOBA, 99.5\%) was used as the solvent for dispersing LSMO and MWCNTs into the PDMS matrix. All chemicals and materials were used as received without further purification.\\ 
\subsection{Synthesis of La$_{0.7}$Sr$_{0.3}$MnO$_3$}
La$_{0.7}$Sr$_{0.3}$MnO$_3$ (LSMO) was synthesized using the conventional solid-state reaction technique. To obtain the desired stoichiometric composition, lanthanum oxide (La$_2$O$_3$) was first subjected to heat treatment at $900\,^{\circ}\mathrm{C}$ for 6~hours, followed by holding at $300\,^{\circ}\mathrm{C}$ for 10~hours in an alumina crucible. This preheating process was carried out to ensure the removal of absorbed moisture and to make La$_2$O$_3$ anhydrous. Subsequently, La$_2$O$_3$, SrCO$_3$, and MnO$_2$ were weighed according to the required stoichiometric ratio and the mixture was thoroughly ground and homogenized in an ethanol medium using an agate mortar and pestle for approximately 3.5~hours. The resulting homogeneous powder mixture was then calcined in a tubular furnace (Ants-I) at $1250\,^{\circ}\mathrm{C}$ for 20~hours to obtain polycrystalline La$_{0.7}$Sr$_{0.3}$MnO$_3$ with the desired nominal composition.\\
\subsection{Fabrication of PDMS/La$_{0.7}$Sr$_{0.3}$MnO$_3$/MWCNT Film}
\begin{figure}
    \centering
    \includegraphics[width=0.48\textwidth]{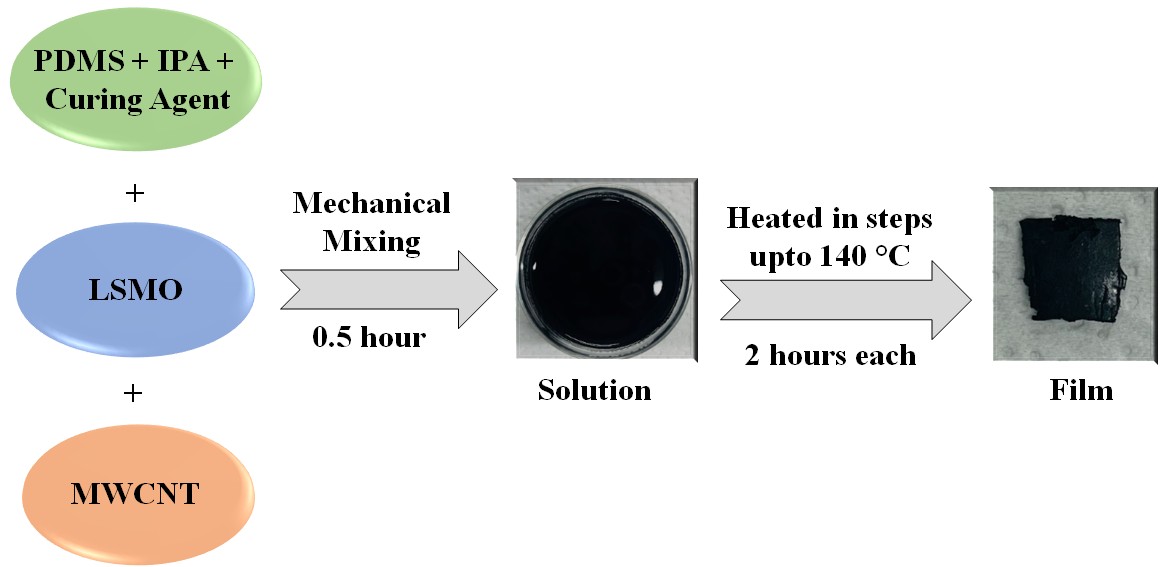}
    \caption{Schematic representation of the preparation procedure for the PDMS/LSMO/MWCNT composite thin film.}
    \label{fig1}
\end{figure}

The schematic representation of the fabrication procedure is shown in Fig.~\ref{fig1}. The nanocomposite thin films were fabricated using a solution mixing method. For the preparation of PDMS/LSMO/MWCNT nanocomposites, PDMS was first dissolved in isopropyl alcohol (IPA) and magnetically stirred for 30~minutes to obtain a homogeneous solution. Subsequently, LSMO and MWCNT were added to the solution in concentrations of 45~wt\% and 5~wt\%, respectively. Subsequently, the PDMS curing agent was added in a 10: 1 base-to-curing agent ratio and the resulting mixture was stirred mechanically using a glass rod for approximately 30~minutes to ensure uniform dispersion of the filler materials throughout the polymer matrix. The prepared solution was then poured into a petri dish and placed inside a vacuum desiccator to remove trapped air bubbles. The degassed mixture was subsequently subjected to sequential thermal curing in a hot air oven at $70\,^{\circ}\mathrm{C}$ for 3~hours, followed by $80\,^{\circ}\mathrm{C}$ for 2~hours, $90\,^{\circ}\mathrm{C}$ for 2~hours, $100\,^{\circ}\mathrm{C}$ for 2~hours, $120\,^{\circ}\mathrm{C}$ for 2~hours, and finally $140\,^{\circ}\mathrm{C}$ for 2~hours. After completion of the curing process, the obtained nanocomposite thin films were carefully peeled off from the petri dish for further characterization and analysis.\\
\subsection{Characterization Details}
The thickness of the fabricated thin films was measured using a digital Vernier caliper (Zhart) with an accuracy of 0.01~mm. For each sample, the thickness was measured at a minimum of five different locations, and the average value was calculated and recorded. The surface morphology and elemental analysis of the samples were investigated using a field-emission scanning electron microscope (FESEM) (CARL ZEISS, GEMINI SEM 500 KMAT model). Atomic force microscopy (AFM) (Bruker) was employed to investigate the surface morphology and surface roughness of the thin films. The AFM measurements were carried out in non-contact mode using an RTESP300-125 probe tip. The surface roughness values were obtained from a scan area of $10~\mu\mathrm{m} \times 10~\mu\mathrm{m}$. X-ray diffraction (XRD) patterns were recorded using a Bruker D8 Advance A25 powder X-ray diffractometer with Cu--K$\alpha$ radiation of wavelength $\lambda = 1.5406$~\AA. The measurements were carried out over a $2\theta$ range of $5^\circ$ to $90^\circ$ at room temperature. The lattice parameters for LSMO were determined through Rietveld refinement analysis. Fourier transform infrared (FTIR) spectroscopy measurements were carried out using an FTIR--ATR spectrometer (PerkinElmer, Singapore) in the wavenumber range of 500--4000~cm$^{-1}$ under attenuated total reflection (ATR) mode. The measurements were performed to investigate the chemical bonding and to confirm the presence of PDMS in the material. Raman spectroscopy measurements were carried out using a confocal Raman microscope (WITec alpha300R). The Raman analysis was performed to investigate and confirm the presence of MWCNTs in the composite thin films. Thermogravimetric analysis (TGA) was carried out using a NETZSCH TG 209F3 instrument from room temperature to $900\,^{\circ}\mathrm{C}$ at a heating rate of $10\,^{\circ}\mathrm{C\,min^{-1}}$ under a nitrogen atmosphere. The analysis was performed to investigate the thermal degradation behavior of the fabricated thin films.\\
The temperature-dependent resistivity measurements were carried out using the standard four-probe method using a source measure unit (SMU, KEITHLEY 2450) connected through pressure probes. The measurements were performed from room temperature up to $170\,^{\circ}\mathrm{C}$. The magnetoresistance (MR) properties of the fabricated thin films were investigated using a Magnetoresistance Research Model (MRX-RM). During the measurements, the sample plane was positioned perpendicular to the applied magnetic field, and the magnetic field was varied from 0~T to 0.6~T at room temperature. The mechanical properties of the fabricated thin films were evaluated using a universal testing machine (UTM) (MTS servo-hydraulic UTM with a 100~kN load cell and Instron 5982). The films were cut into standardized test specimens according to the ISO 527-3 standard. Tensile tests were performed at a crosshead speed of 1~mm/min$^{-1}$.\\
\section{Results and Discussion}
\subsection{FESEM Analysis}
\begin{figure*}[t]
    \centering
    \includegraphics[width=0.9\textwidth]{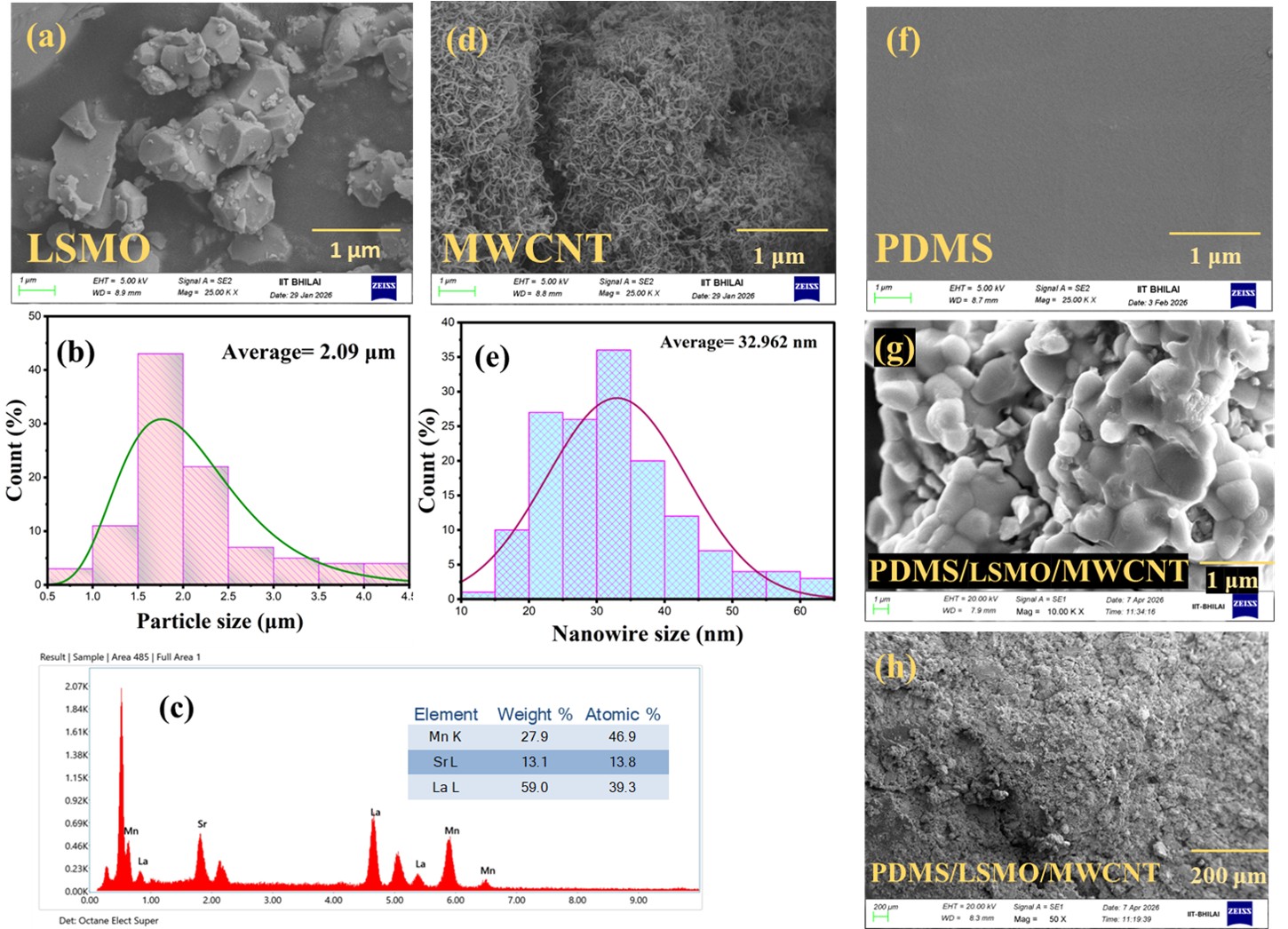}
    \caption{(a) shows the FESEM image of solid-state synthesized La$_{0.7}$Sr$_{0.3}$MnO$_3$ (LSMO), illustrating its surface morphology. (b) presents the histogram of the particle size distribution of the LSMO particles. (c) shows the elemental mapping of the synthesized LSMO ceramic powder. (d) depicts the FESEM image of multi-walled carbon nanotubes (MWCNTs). (e) shows the histogram corresponding to the diameter distribution of the MWCNTs. (f) presents the FESEM image of the surface morphology of pristine PDMS at higher magnification, whereas (g--h) show the surface morphology of the PDMS/LSMO/MWCNT composite thin film at different scanning scales.}
    \label{fig2}
\end{figure*}

Figure~\ref{fig2}(a) shows the FESEM image of solid-state synthesized La$_{0.7}$Sr$_{0.3}$MnO$_3$ (LSMO), illustrating the surface morphology of the synthesized particles. The average particle size was analyzed using \textit{ImageJ} software and was found to be approximately $2.09~\mu$m, as shown in Figure~\ref{fig2}(b). The elemental mapping analysis presented in Figure~\ref{fig2}(c) confirms the presence of lanthanum (La), strontium (Sr), and manganese (Mn) in the synthesized sample. Furthermore, the elemental ratio of La:Sr:Mn was found to be close to the nominal stoichiometric composition of 0.7:0.3:1, indicating a successful synthesis of the desired LSMO phase.\\
The loosely agglomerated morphology of MWCNT powder dispersed in the PDMS polymer matrix in the presence of IPA was characterized using FESEM, as shown in Figure~\ref{fig2}(d). The average nanotube diameter was determined to be approximately $32.962$~nm, as presented in Figure~\ref{fig2}(e).\\
The surface morphology of the pristine PDMS film is shown in Figure~\ref{fig2}(f), where a smooth and uniform surface is observed. However, upon incorporation of LSMO and MWCNT fillers into the PDMS matrix, the smoothness of the polymer surface is significantly altered, as evident from Figure~\ref{fig2}(g). In the composite film, the nearly cubic morphology of the LSMO particles is not distinctly visible, suggesting that the particles are embedded within the polymer matrix. The embedded ceramic particles appear as protruded regions distributed across the composite surface, as observed in Figure~\ref{fig2}(g--h).\\
\subsection{AFM Analysis}
\begin{figure}
    \centering
    \includegraphics[width=0.49\textwidth]{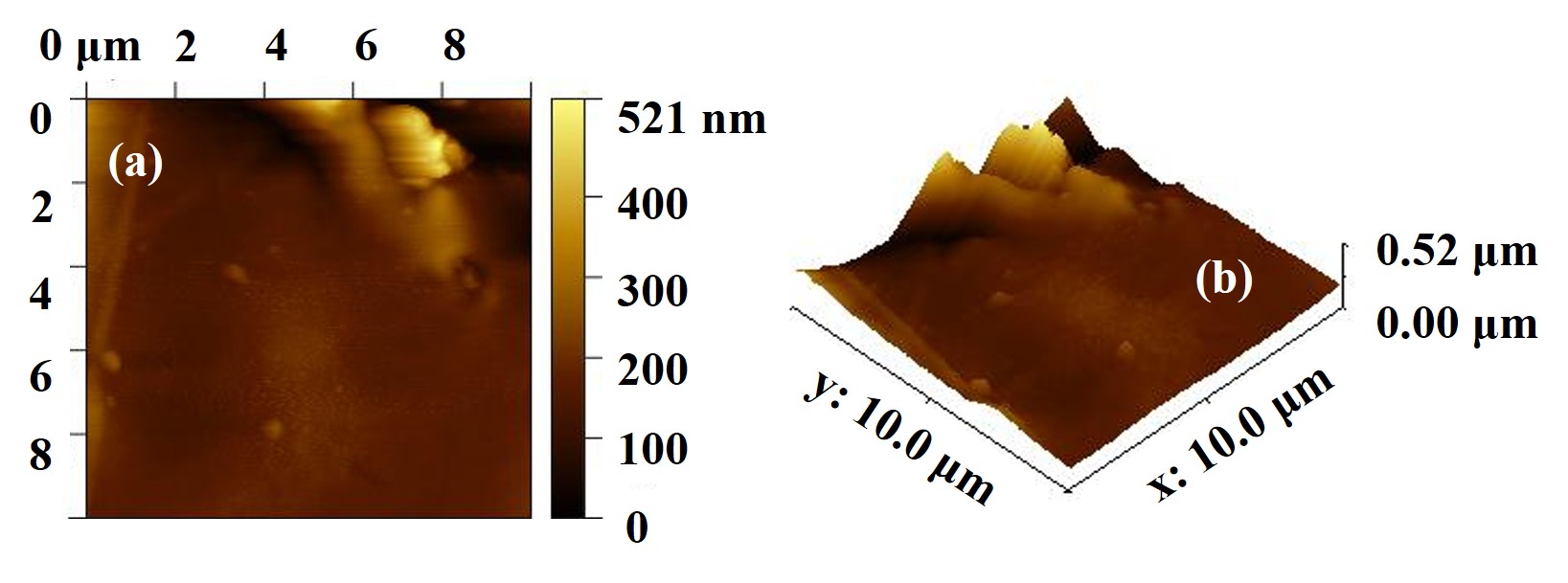}
    \caption{(a) Two-dimensional (2D) AFM surface morphology image of the PDMS/LSMO/MWCNT composite thin film. (b) Three-dimensional (3D) AFM surface morphology image of the PDMS/LSMO/MWCNT composite thin film.}
    \label{fig3}
\end{figure}

The surface morphology of the composite thin film was investigated using atomic force microscopy (AFM) and is shown in Fig.~\ref{fig3}. The surface roughness measurements were carried out over a scan area of $10~\mu\mathrm{m} \times 10~\mu\mathrm{m}$. The average surface roughness of the film was found to be approximately $254.7$~nm, as analyzed using \textit{Gwyddion} software\cite{zhao2019quantitative}, and the root mean square (RMS) roughness value was determined to be $22.15$~nm.\\
The thickness of the fabricated thin film was measured using a digital caliper and was found to be approximately $0.85$~mm. Compared to the overall thickness of the film, the measured surface roughness is negligible, indicating that the fabricated films possess comparatively uniform and smooth surfaces. The detailed roughness parameters obtained from the AFM analysis are summarized in Table~\ref{tab:afm_roughness}.\\
\begin{table}[htbp]
\centering
\caption{Roughness Parameters of PDMS/LSMO/MWCNT Film using AFM}
\label{tab:afm_roughness}
\renewcommand{\arraystretch}{1.2}
\begin{tabular}{|l|c|}
\hline
\textbf{Parameters} & \textbf{Value} \\
\hline
Average Roughness & 254.7 nm \\
\hline
RMS Roughness (Sq) & 22.15 nm \\
\hline
RMS (Grain-Wise) & 22.15 nm \\
\hline
Mean Roughness (Sa) & 12.41 nm \\
\hline
Skew (Ssk) & -0.8317 \\
\hline
Excess Kurtosis & 21.27 \\
\hline
\end{tabular}
\end{table}
Few important observations that can be emphasized here are, the composite film exhibits a hierarchically structured surface, characterized by multi-scale roughness arising from the dispersion of LSMO particles and the percolated MWCNT network within the PDMS matrix. Negative skewness indicates a valley-dominated morphology, which enhances conformability and mechanical adaptability when interfaced with soft biological tissues, a critical requirement for e‑skin applications. The exceptionally high kurtosis value suggests the presence of sharp and localized surface features, indicative of heterogeneous filler distribution and percolation pathways, which can enhance local field concentration and improve sensing performance. The moderate RMS roughness (22~nm) ensures an optimal balance between surface uniformity and functional irregularity, enabling stable electrical response while enhancing sensitivity to external stimuli.\\
\subsection{XRD Analysis}
\begin{figure*}
    \centering
    \includegraphics[width=0.9\textwidth]{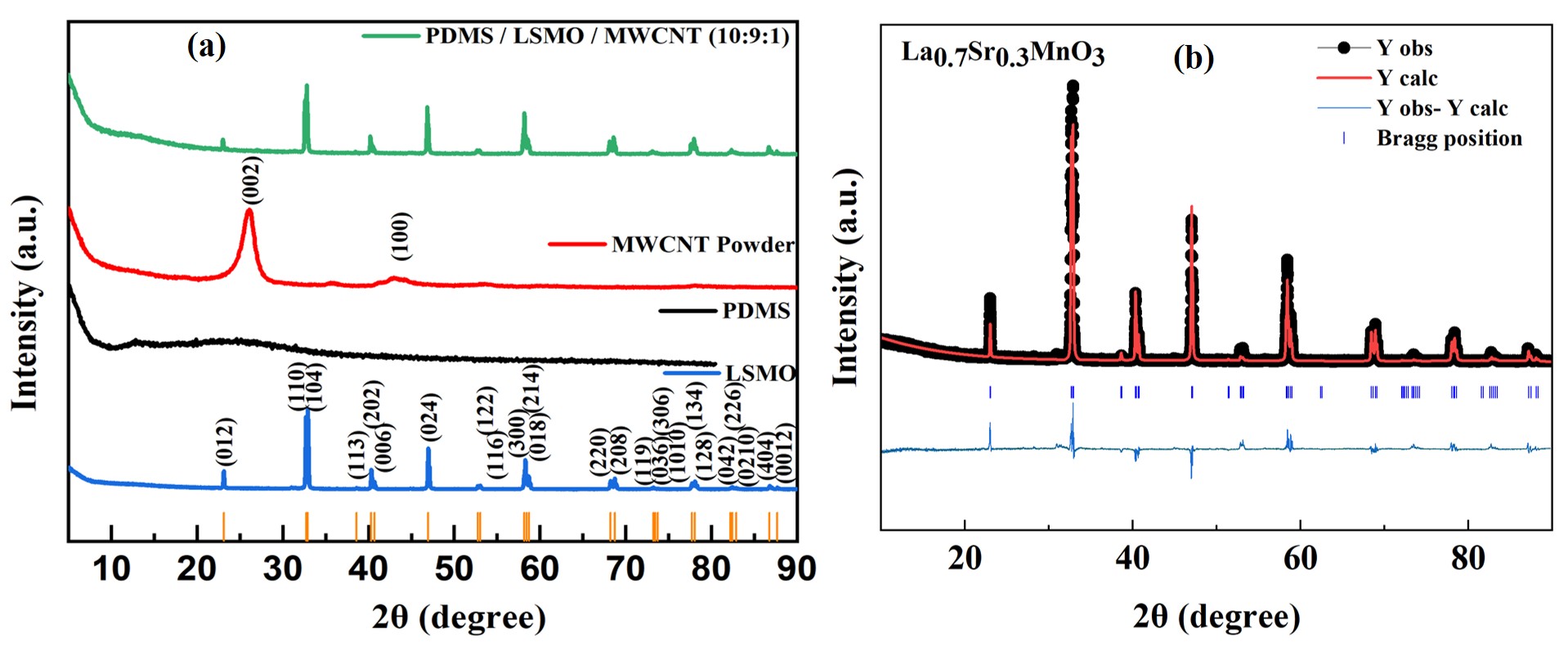}
    \caption{(a) XRD Pattern of LSMO, PDMS, MWCNT and PDMS/LSMO/MWCNT Thin Film (b) Rietveld Refinement of LSMO.}
    \label{fig4}
\end{figure*}
The X-ray diffraction (XRD) patterns of La$_{0.7}$Sr$_{0.3}$MnO$_3$ (LSMO), MWCNT, PDMS, and the PDMS/LSMO/MWCNT composite thin film are shown in Figure~\ref{fig4}(a). The diffraction pattern confirms the formation of a single-phase LSMO compound. The observed Bragg peak positions and their corresponding Miller indices are indexed and matched with the JCPDS file No.~96-722-2424, indicating that the crystal structure of LSMO belongs to the hexagonal crystal system with the rhombohedral space group $R\bar{3}c$ (space group number 167). These results are further supported by the Rietveld refinement analysis discussed later.

The diffraction peaks corresponding to MWCNTs were observed at the (002) and (100) planes with 2$\theta$ values around $26^\circ$ and $43^\circ$, respectively, which are in agreement with the previously reported literature values~\cite{nie2015preparation}. No prominent diffraction peaks were observed for PDMS due to its predominantly amorphous nature. In the XRD pattern of the composite thin film, the characteristic peaks of LSMO are clearly visible, whereas the diffraction peaks corresponding to MWCNTs are not distinguishable, possibly due to their lower concentration and overlap with the broad amorphous background of PDMS.

The powder XRD pattern of LSMO was further analyzed by Rietveld refinement using the \textit{FullProf}$^{\mathrm{TM}}$ software package\cite{rietveld1969profile}. The refinement was carried out considering the $R\bar{3}c$ space group, and the refined pattern is shown in Figure~\ref{fig4}(b). The refined lattice parameters were obtained as $a = 5.478920$~\AA, $b = 5.478920$~\AA, $c = 13.297576$~\AA, $\alpha = 90.000000^\circ$, $\beta = 90.000000^\circ$, and $\gamma = 120.000000^\circ$. The refinement quality parameters were found to be $R_p = 33.7$, $R_{wp} = 26.7$, $R_{exp} = 11.70$, and $\chi^2 = 5.21$.\\

\subsection{Spectroscopic Analysis}
\begin{figure*}
    \centering
    \includegraphics[width=0.95\textwidth]{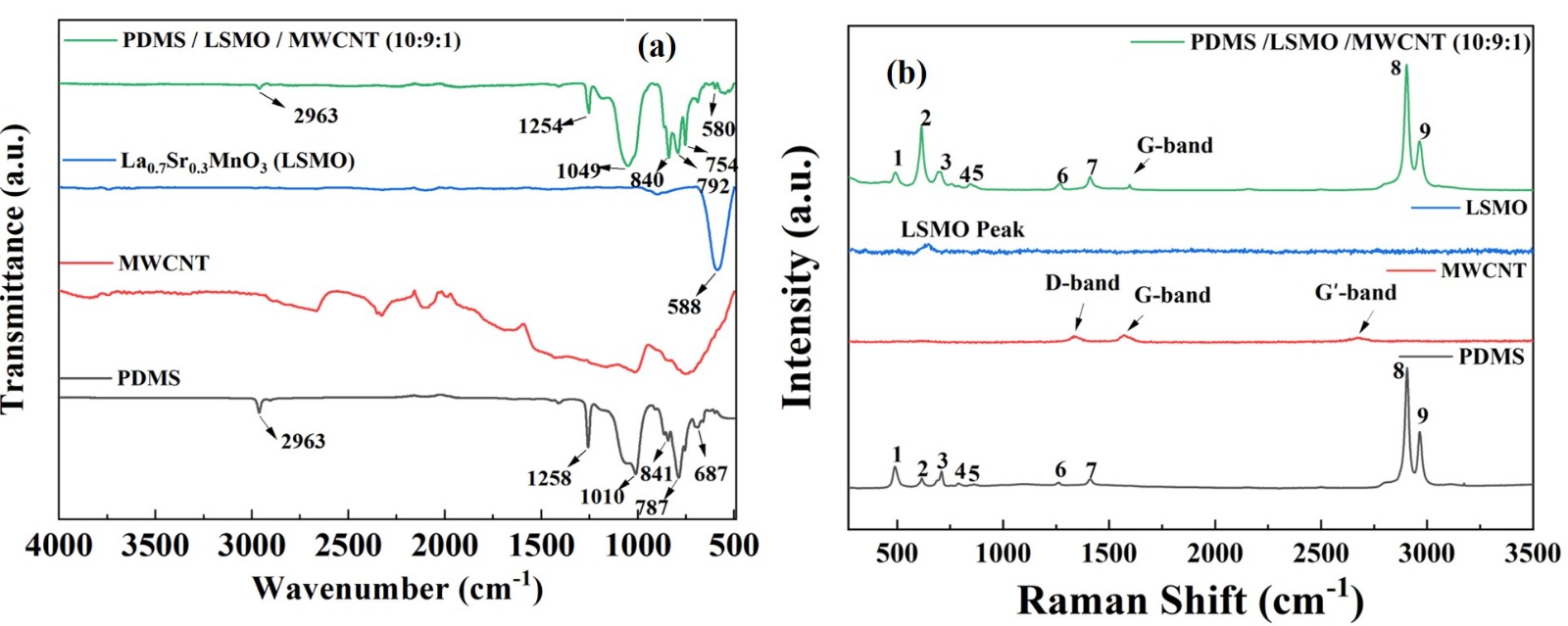}
    \caption{ (a) FTIR Spectra of PDMS, MWCNT, LSMO, and PDMS/LSMO/MWCNT Thin film, (b) Raman Spectrum of PDMS, MWCNT, LSMO, and PDMS/LSMO/MWCNT thin film.}
    \label{fig5}
\end{figure*}
To verify the possible interactions among the constituents of the thin films, Fourier transform infrared (FTIR) spectra of the pristine components, namely LSMO, PDMS, and MWCNT, together with the PDMS/LSMO/MWCNT composite thin film, are presented in Figure~\ref{fig5}. The FTIR spectra of pristine PDMS and the composite thin film exhibit very similar features, indicating that the polymer matrix dominates the infrared response of the composite. The characteristic absorption peaks corresponding to MWCNTs are not distinctly visible in the composite film. This may be attributed to the relatively lower concentration of MWCNTs and the masking effect caused by the strong absorption bands of PDMS. The observed absorption peaks and their corresponding functional groups or vibrational bonds are summarized in Table~\ref{tab:FTIR}, and are in good agreement with the values reported earlier in the literature~\cite{thorat2013functionalization,razi2018magnetoresistance,wu2021study,hamouni2019alcohol}. The FTIR observations that arise due to dipole moment changes could be interpreted as follows: the 1010 $\mathrm{cm}^{-1}$ peak of pristine PDMS shift towards longer wavelength side in the composite film, it may be due to the interaction between PDMS and LSMO oxide interfaces. Being it a physical adsorption, it may hinder the polymer chain mobility. The reduction in the intensity of the peak at 2963 $\mathrm{cm}^{-1}$ with respect to the pristine PDMS could be due to the steric hinderances due to the fillers. The LSMO's low frequency region overlaps with PDMS peaks, which is reflected in the composite film as a small shift. This could be attributed to the strain transfer from the polymer matrix and the interfacial lattice distortions in LSMO.\\
\begin{table*}[htbp]
\centering
\caption{The wavenumber (cm$^{-1}$) corresponding functional group and their vibrational mode for the spectra obtained from PDMS and LSMO samples.}
\label{tab:FTIR}
\renewcommand{\arraystretch}{1.2}
\begin{tabular}{|c|c|c|l|}
\hline
\textbf{Component} & \textbf{Wavenumber (cm$^{-1}$)} & \textbf{Functional Group / Bond} & \textbf{Vibrational Mode} \\
\hline
PDMS & 2963 & Si-CH$_3$ & Symmetrical stretching of -CH$_3$ \\
\hline
PDMS & 1258 & -CH$_3$ & Rocking \\
\hline
PDMS & 1010 & Si-O-Si & Symmetrical stretching of Si-O \\
\hline
PDMS & 841 & Si-C-H & Stretching of Si-C-H \\
\hline
PDMS & 687 & Si-O-Si & Bending vibrations of Si-O \\
\hline
LSMO & 588 & Mn-O & Stretching (lattice vibration) of Mn-O-Mn \\
\hline
\end{tabular}
\end{table*}
To further confirm the incorporation of MWCNTs into the composite thin film, Raman spectroscopy measurements were carried out at room temperature for pristine PDMS film, LSMO powder, MWCNT powder, and the PDMS/LSMO/MWCNT composite film, as shown in Figure~\ref{fig5}(b). The observed Raman shifts and their corresponding vibrational modes are listed in Table~\ref{tab:raman_modes}. The obtained peak positions are in close agreement with the values reported for bulk PDMS~\cite{bae2005chemical}, LSMO powder~\cite{kumari2020structural}, and MWCNTs~\cite{zeng2015encapsulating,yashiro2016evaluation,aoki2023raman}. The Raman spectra of pristine PDMS and the composite film show nearly identical peak positions, although a reduction in the intensity of peak number (2) is observed in the composite film. This may indicate an improvement in the interpretation of interfacial and crystalline contributions between LSMO and PDMS. This LSMO phonon shift may affect the magnetic/ electric behavior leads to variations in electrical conductivity / magnetoresistive response.\\
The characteristic Raman peaks of LSMO are not distinctly visible in the composite spectrum, which may be due to the dominant Raman response of PDMS in the same spectral region. For MWCNTs, the characteristic D-band and G$'$-band are not clearly observed in the composite film. However, the G-band located at $1578~\mathrm{cm}^{-1}$ in pristine MWCNTs is observed in the composite film with a slight shift to $1601~\mathrm{cm}^{-1}$, denoted as the G-band.\cite{antunes2006comparative} The shift in the G-band position may originate from interfacial interactions between MWCNTs and the PDMS/LSMO matrix that lead to  charge transfer between CNT and LSMO. The presence of this shifted G-band confirms the successful incorporation of MWCNTs into the composite thin film. The peak shifts and intensity distributions in the composite indicated the improved mechanical coupling between the constituents which is a good indication for strain/ pressure sensing\cite{avramenko2022role}.\\
\begin{table*}[htbp]
\centering
\caption{Raman shift (cm$^{-1}$) corresponding vibrational modes observed for PDMS, MWCNT and LSMO components.}
\label{tab:raman_modes}
\renewcommand{\arraystretch}{1.2}
\begin{tabular}{|c|c|c|l|}
\hline
\textbf{Symbol} & \textbf{Component} & \textbf{Raman Shift (cm$^{-1}$)} & \textbf{Vibration Mode} \\
\hline
1 & PDMS & 491  & Si--O--Si symmetric stretching \\
\hline
2 & PDMS & 621  & Si--CH$_3$ symmetric rocking \\
\hline
3 & PDMS & 713  & Si--C symmetric stretching \\
\hline
4 & PDMS & 802  & CH$_3$ asymmetric rocking + Si--C asymmetric stretching \\
\hline
5 & PDMS & 857  & CH$_3$ symmetric rocking \\
\hline
6 & PDMS & 1272 & CH$_3$ symmetric bending \\
\hline
7 & PDMS & 1416 & CH$_3$ asymmetric bending \\
\hline
8 & PDMS & 2900 & CH$_3$ symmetric stretching \\
\hline
9 & PDMS & 2965 & CH$_3$ asymmetric stretching \\
\hline
D-band & MWCNT & 1345 & Disordered structure or sp$^3$ hybridized carbons \\
\hline
G-band & MWCNT & 1578 & Splitting of the E$_{2g}$ stretching mode of graphite \\
\hline
G$'$-band & MWCNT & 2680 & 2D band \\
\hline
G-Band & MWCNT & 1601 & Shifted G-band \\
\hline
LSMO Peak & LSMO & 641 & Presence of MnO$_6$ octahedra \\
\hline

\end{tabular}
\end{table*}

\subsection{Thermal Analysis}
\begin{figure}
    \centering
    \includegraphics[width=0.45\textwidth]{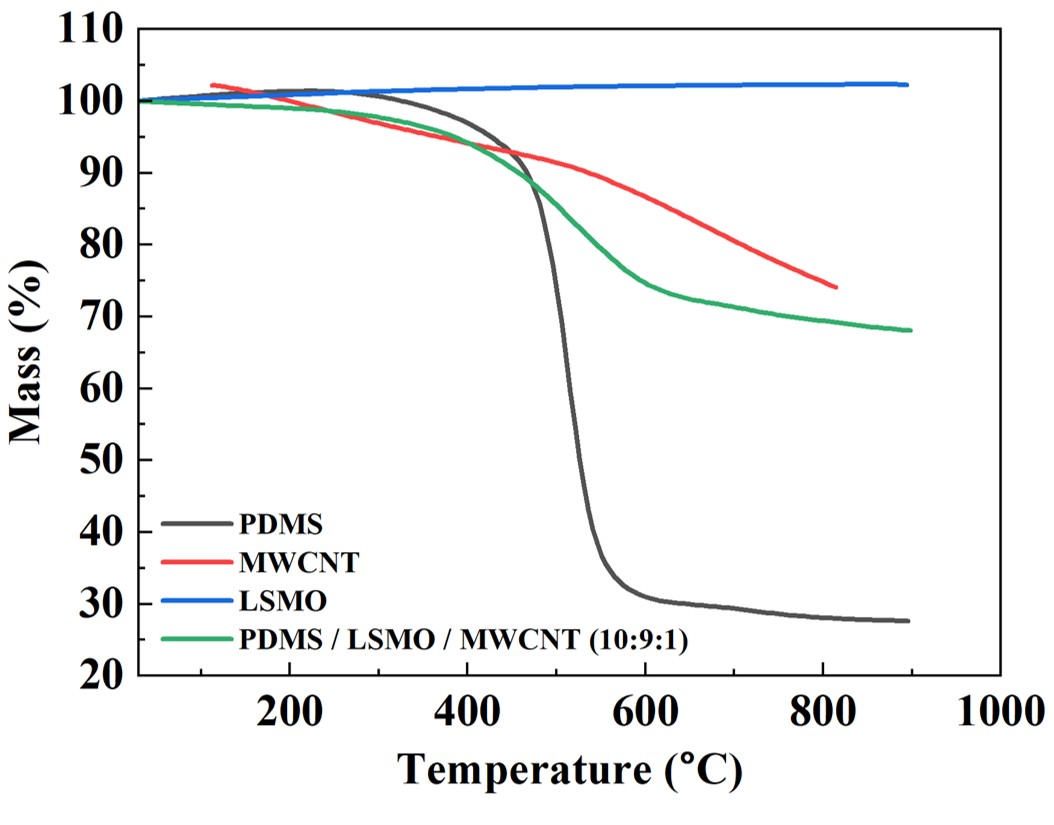}
    \caption{Thermogravimetric analysis (TGA) curves of pristine PDMS film, MWCNT, LSMO, and PDMS/LSMO/MWCNT (10:9:1) composite thin film measured under N$_2$ atmosphere at a heating rate of $10\,^{\circ}\mathrm{C\,min^{-1}}$.}
    \label{fig6}
\end{figure}
Thermogravimetric analysis (TGA) thermograms were recorded for pristine PDMS film, MWCNTs, LSMO, and the PDMS/LSMO/MWCNT composite thin film. As observed from Figure~\ref{fig6}, the thermal degradation of pristine PDMS begins at approximately $200\,^{\circ}\mathrm{C}$ and continues up to nearly $600\,^{\circ}\mathrm{C}$, after which a plateau region is observed, indicating the completion of major thermal decomposition.\cite{camino2001polydimethylsiloxane}\\
In the case of MWCNTs, no sharp decrease in mass is observed as a results of thermal degradation within the investigated temperature range. However, an overall mass loss of nearly 20\% is evident from Figure~\ref{fig6}. Similarly, LSMO does not exhibit any noticeable thermal degradation. This enhanced thermal stability can be attributed to the high-temperature calcination involved during the solid-state synthesis process of LSMO, which already exposes the material to elevated temperatures.\\
For the PDMS/LSMO/MWCNT composite thin film, the onset of thermal degradation occurs approximately at $400\,^{\circ}\mathrm{C}$. In comparison, pristine PDMS begins to degrade near $200\,^{\circ}\mathrm{C}$. Therefore, incorporation of thermally stable LSMO and MWCNT fillers significantly enhances the thermal stability of the composite thin film, increasing its effective operating temperature range to nearly $400\,^{\circ}\mathrm{C}$, as evidenced by the higher residual mass of the composite film compared to pristine PDMS at elevated temperatures. This indicates that the developed composite film can safely operate within a considerably higher temperature range than pristine PDMS, ensuring reliability under thermal fluctuations and prolonged usage. Interestingly, the virgin PDMS shows a sharp degradation while the composite shows broadened, gradual decomposition, indicating a strong polymer–filler interaction and restricted polymer chain mobility due to the presence of LSMO particles and MWCNT network. The negligible mass loss observed for LSMO confirms its role as a thermally inert component, contributing significantly to the enhanced thermal robustness of the composite system. The heat resistant and thermally conductive networks of MWCNT may also suppress the composite volatilization.\\
\subsection{Electrical and Magnetic Characterization}
\begin{figure*}
    \centering
    \includegraphics[width=0.8\textwidth]{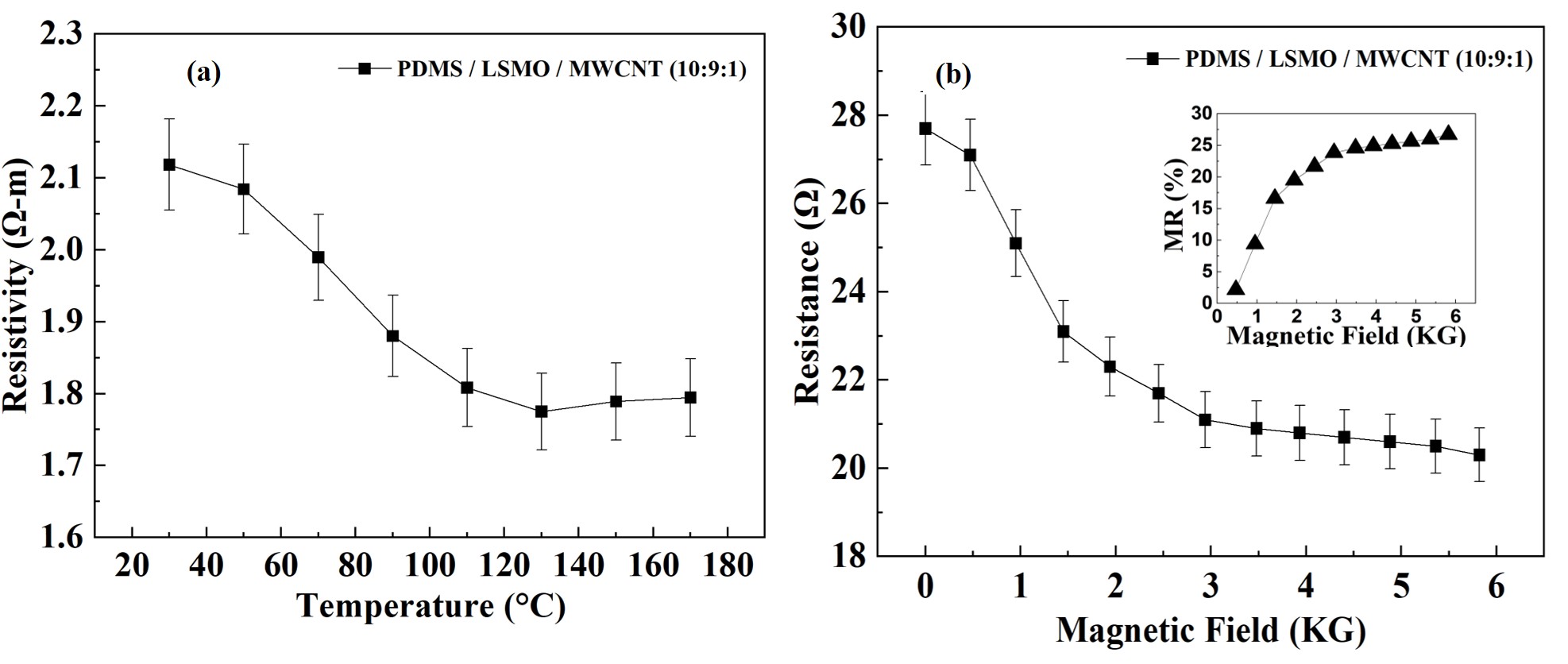}
    \caption{(a) Temperature-dependent variation in the electrical resistivity of the PDMS/LSMO/MWCNT composite thin film. (b) Magnetic field-dependent variation in the electrical resistance of the PDMS/LSMO/MWCNT composite thin film. Inset shows the magnetoresistance (MR) behavior of the PDMS/LSMO/MWCNT composite thin film under an applied magnetic field.}
    \label{fig7}
\end{figure*}

Figure~\ref{fig7}(a) shows the temperature dependence of the electrical resistivity of the PDMS/LSMO/MWCNT composite thin film. Two distinct temperature regions can be observed from the resistivity profile. The first region extends from room temperature ($30\,^{\circ}\mathrm{C}$) to approximately $130\,^{\circ}\mathrm{C}$, while the second region lies between $130\,^{\circ}\mathrm{C}$ and $170\,^{\circ}\mathrm{C}$.
Initially, the electrical resistivity of the composite film was found to be approximately $2.11~\Omega$-m at room temperature. As the temperature increased, the resistivity gradually decreased to nearly $1.77~\Omega$-m up to $130\,^{\circ}\mathrm{C}$. Beyond this temperature, the resistivity exhibited a slight increase from $1.77~\Omega$-m to approximately $1.79~\Omega$-m with a further increase in temperature up to $170\,^{\circ}\mathrm{C}$, which is $\sim$17\% change in resistivity with $\sim$100$^{\circ}$ change in temperature. The observed temperature-dependent variation in resistivity confirms that the fabricated composite thin film exhibits temperature sensing characteristics, thereby fulfilling one of the primary objectives of the present study. Furthermore, The observed decrease in resistivity with increasing temperature confirms the semiconducting behavior of the composite, governed by thermally activated charge transport through the MWCNT percolation network and interfacial regions\cite{kumar2020graphene, kumar2023magnetic, pandey2024monolithic}.\\
To test another functionality of this film, we studied the change in resistance under applied magnetic field. Figure~\ref{fig7}(b) illustrates the variation of electrical resistance of the composite film under the application of an external magnetic field. It is observed that the electrical resistance decreases with increasing magnetic field strength. This behavior can be attributed to the contribution of the LSMO phase, which is the magnetic component present in the composite film.\\
\begin{figure*}
    \centering
    \includegraphics[width=0.9\textwidth]{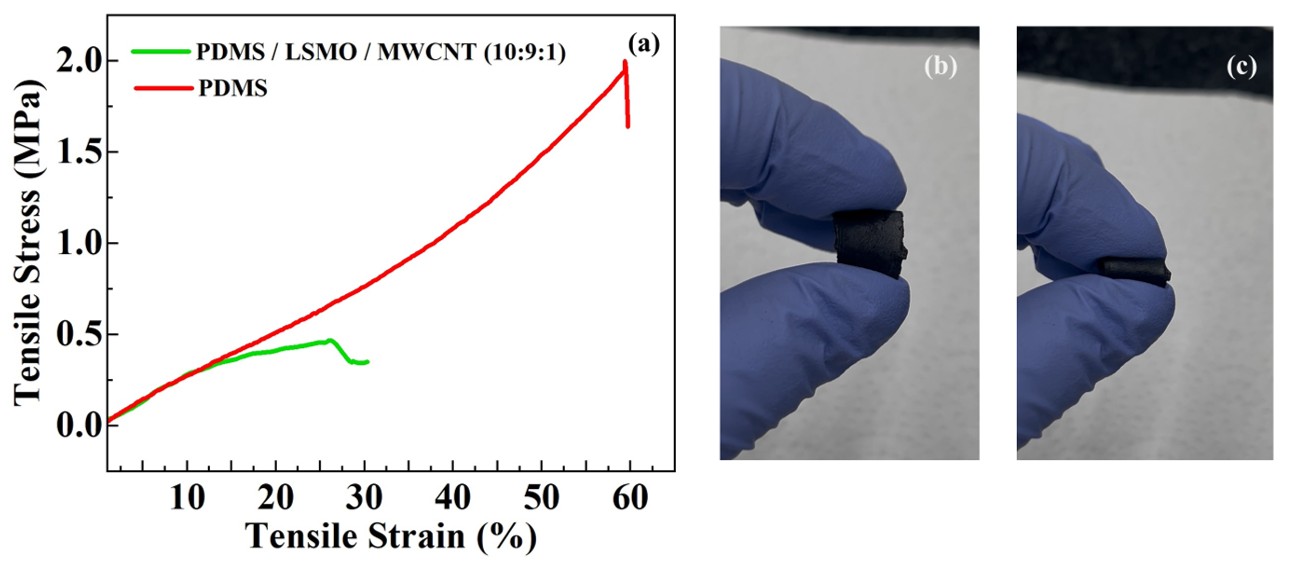}
    \caption{(a)Tensile stress--strain behavior of the PDMS/LSMO/MWCNT composite thin film obtained from uniaxial tensile testing. (b-c) shows the flexible nature of the film.}
    \label{fig8}
\end{figure*}
Inset of Fig.~\ref{fig7}(b) presents the magnetoresistance (MR) versus magnetic field plot, which was used to understand the magnetic field-dependent transport behavior of the composite thin film. The magnetoresistance percentage was calculated using the following relation:
\begin{equation}
\mathrm{MR}~(\%) = \left( \frac{R(0) - R(H)}{R(0)} \right) \times 100
\label{eq:MR}
\end{equation}

where, R(0) and R(H) are the Resistance in the absence and presence of magnetic field, respectively, and MR is called magnetoresistance. The composite exhibits pronounced negative magnetoresistance reaching a value of nearly 27\%, attributed to spin-polarized transport within the LSMO component, where magnetic field-induced alignment of spins reduces electron scattering and enhances conductivity\cite{salamon2001physics,kumar2018magnetothermopower}. The sharp reduction in resistance at low magnetic fields indicates high sensitivity in the low-field regime, which is desirable for wearable magnetic sensing applications. The saturation of magnetoresistance at higher magnetic fields suggests a complete alignment of magnetic domains, beyond which additional field strength does not significantly influence electron transport. In general, the composite exhibits stable and responsive electrical behavior under thermal and magnetic stimuli, making it a promising candidate for multifunctional electronic skin applications.\\

\subsection{Mechanical Properties}
The mechanical properties of the composite thin film were evaluated using the stress--strain behavior obtained from tensile testing performed on a universal testing machine (UTM). Figure~\ref{fig8}(a) shows the stress--strain curve of the fabricated composite film along with PDMS. PDMS possesses a tensile strength of $2.0$~MPa and a strain of $\sim 60\%$. It is observed that the maximum load sustained by the composite film is approximately $2.78$~N, corresponding to a tensile strength of $0.49$~MPa and a strain of $26.16\%$. Thus the composite film exhibits pronounced elastomeric behavior with elongation up to $\sim$26\%, confirming that the PDMS matrix effectively preserves flexibility even after incorporation of LSMO and MWCNT fillers. The gradual enhancement of tensile strength compared to pristine PDMS suggests effective load transfer between the polymer matrix and the dispersed LSMO particles and MWCNT network, indicating strong interfacial interactions~\cite{sidhu2025development}. The absence of abrupt fracture and the presence of a gradual stress decline beyond the maximum stress indicate ductile failure behavior, which is desirable for wearable and flexible electronic applications. Furthermore, the fracture strain of the composite film was found to be approximately $30.30\%$, indicating that the fabricated thin film possesses appreciable flexibility and stretchability suitable for electronic skin applications. From the initial linear region (0--5\% strain) the Young's modulus is approximately found to be 2 MPa which is comparable to the human skin~\cite{wahlsten2023multiscale}. From the observed ultimate tensile strength ($\sim$46\%) and elongation at break ($\sim$26\%), using the triangular approximation, toughness is calculated to be 0.056 MJ/m$^3$ which is sufficient for repeated deformation cycles~\cite{zhao2017mechanisms}. The bending images (Fig.~\ref{fig8}(b-c)) demonstrate excellent flexibility and mechanical compliance, with no visible structural damage, confirming the suitability of the composite for conformal and wearable applications. The smooth stress increase indicates a good dispersion of fillers, and no sudden drop of stress indicates a strong interfacial bonding between the matrix and particulate component. To achieve this feat, the MWCNT network plays a critical role in enhancing mechanical resilience by redistributing stress and bridging microcracks during deformation.\\

\section{Conclusions}
PDMS/LSMO/MWCNT composite thin films were fabricated through a solution casting technique. XRD confirms the presence of each phase and elemental analysis established the presence of constituent elements in the composite film. The surface roughness analysis (from AFM measurements) revealed a smooth and uniform surface of the film. Thermogravimetric analysis demonstrated the enhanced thermal stability of the film. Notably, the composite retains $\sim$ 70\% of its mass at elevated temperatures, substantially higher than that of pure PDMS, validating the effective incorporation of thermally stable components. These results highlight the suitability of the film for $e-skin$ applications, where thermal robustness and long-term operational stability are critical. The FTIR investigation confirmed the presence of PDMS in the composite film, while Raman spectroscopy successfully verified the incorporation of MWCNTs through the observation of characteristic Raman bands. Temperature-dependent resistivity measurement depicts the thermoreceptor functionality of human skin. Furthermore, the electrical resistance also decreases with applied magnetic field, confirming the magnetic field-responsive behavior of the film owing to the presence of magnetic LSMO phase. Mechanical characterization revealed that the maximum load sustained by the film is approximately $2.78$~N, indicating an appreciable mechanical flexibility suitable for e-skin applications. The absence of abrupt fracture and the presence of a gradual stress decline beyond the maximum stress indicate ductile failure behavior with a stretchability of $\sim26\%$, which is desirable for wearable and flexible electronic applications. These combined observations suggest the formation of a percolative, mechanically coupled, and electronically interactive composite system, which is critical for multifunctional sensing applications and arises from the synergistic interaction between the conductive MWCNT network and the magnetically active LSMO phase embedded within the flexible PDMS matrix. In conclusion, thermally stable, electrically responsive, magnetically active and flexible material is tested and reported for next generation $e-skin$ applications.\\

\bibliographystyle{apsrev4-2}
\bibliography{ref}
\end{document}